\documentclass[aps,prl,reprint,amsmath,amssymb,longbibliography,nofootinbib,floatfix]{revtex4-2}
\usepackage[T1]{fontenc}
\usepackage{graphicx,bm,booktabs}
\usepackage[colorlinks=true,allcolors=blue]{hyperref}
\newcommand{\tr}{\operatorname{tr}}
\newcommand{\Fretain}{F_R}
\newcommand{\Fglobal}{F_{\mathrm{full}}}
\begin{document}
\title{Optimal local recovery cannot alter critical orthogonality exponents in quantum spin environments}

\author{Zhixuan Zhao}
\author{Jun Li}
\email{ljcj007@ysu.edu.cn}
\affiliation{
	State Key Laboratory of Metastable Materials Science and Technology, Hebei Key Laboratory of Microstructural Material Physics, School of Science, Yanshan University, Qinhuangdao, 066004, China.}

\date{\today}
\begin{abstract}
	Interference disappears not when a probe falters, but when its
	environment remembers which state the probe occupied. When the two
	probe states correlate with distinguishable ground states of the
	environment, fringes fade; at a critical point, even a local
	perturbation can drive the two ground states toward orthogonality as
	the system grows. A natural remedy is quantum erasure: if an arbitrary
	unitary, controlled by the probe, may act on a fixed number of
	environmental spins, how much interference can be recovered? We give a
	sharp answer. The recovery gain is bounded by local energy scales
	alone. For transverse-field Hamiltonians whose ground-state amplitudes
	are positive in a common spin basis, with uniformly bounded ratios of
	local longitudinal bias to transverse field, optimal recovery can
	strongly enhance visibility yet cannot alter its power-law decay
	exponent, and the bound requires no gap anywhere else in the
	environment. In a connected critical Ising chain, the exponent $1/8$
	survives optimal recovery acting on the two spins adjoining the defect
	bond; an auxiliary-spin model exhibits instead a large finite-size
	enhancement. The orthogonality exponent thus governs the genuinely
	collective part of interference loss, persisting after every locally
	recoverable contribution has been removed.
\end{abstract}
\maketitle

Probe interferometry turns the response of a many-body environment into
a measurable signal. When the two states of a probe prepare different
environment states, their distinguishability reduces its interference
\cite{Englert1996}. For equally weighted coherent branches, the fringe
visibility is the magnitude of the environmental overlap. This connects
interference to a central many-body effect: even a local perturbation can
make two ground states orthogonal as the environment grows
\cite{Anderson1967,Munder2012}. The decay of their overlap with system
size records a collective rearrangement in a single-probe observable.

A weak signal alone, however, does not establish this collective origin.
A few nearby spins can carry an almost perfectly distinguishable record of the
probe and strongly suppress its visibility. Removing that record can
restore interference, the aim of quantum erasure \cite{Scully1982}.
With access to only part of the environment, recovery depends on which
degrees of freedom are available \cite{Miatto2015}. This creates a
physical distinction that the bare overlap does not resolve: how much
suppression can be undone by improving control over the same few spins,
and what loss necessarily persists as the environment grows? In
particular, an orthogonality exponent characterizes the initial overlap;
its relevance to optimally recovered interference requires a further
constraint on local control.

We study deterministic unitary recovery on selected physical spins,
conditioned on the probe state. This operation can align their local
records without changing the uncontrolled part of either branch. Its
largest recovered overlap equals the fidelity between the two states of
that uncontrolled remainder \cite{Uhlmann1976,Jozsa1994}; simply tracing
out the selected spins is a way to calculate this optimum, not a recovery
operation. Reduced-state fidelity has also been used to diagnose quantum
transitions \cite{Paunkovic2008}. Its monotonicity guarantees that optimal
local recovery cannot reduce visibility, but gives no limit on the gain.
Such a limit is needed to decide whether a vanishing initial overlap
remains an obstacle after the best available local intervention.

Here we obtain this limit from local energy scales. For the
positive-ground-state Hamiltonians defined below, transverse spin
fluctuations restrict how distinguishable a local record can become.
The maximum recovery gain is a finite multiplicative factor fixed by
local longitudinal biases and transverse fields, independent of the
environment's total size or gap. Thus an existing orthogonality exponent
survives optimal control of a fixed number of spins with bounded local
ratios. It constrains recovered interference, not only the initial
wavefunction overlap. A connected critical Ising chain shows that keeping
visibility nonzero requires the controlled region to grow. A complementary
auxiliary-spin example exhibits a large recoverable suppression at finite
size. Together they distinguish improving an interference signal from
removing the collective decay that ultimately limits it.

\emph{A Hamiltonian bound on local recovery.---}
The two probe states, labeled $z=\pm$, select two Hamiltonians for the
environment. Partition it into the $k$ spins $M$ available for recovery
and the uncontrolled remainder $R$. In a product basis, $x$ labels $R$
and $y$ labels the $Z$ orientations of $M$. We consider
\begin{equation}
H_z=H_{R,z}-\sum_{\ell\in M}h_\ell X_\ell+V_z(x,y),
\qquad h_\ell>0.
\label{eq:H}
\end{equation}
$X_\ell$ and $Z_\ell$ are Pauli operators on a selected spin.
In the same product basis, $H_{R,z}$ is real symmetric with nonpositive
off-diagonal entries, and $V_z$ is real diagonal. Irreducibility makes each
normalized ground state $\Omega_z(x,y)$ strictly positive. Diagonal
interactions within $M$, spatial disorder, and frustrated diagonal couplings
are allowed; the selected spins flip only through the displayed transverse
fields. Assume the local longitudinal-bias bounds
\begin{equation}
\frac12\left|V_z(x,+_\ell,y')-V_z(x,-_\ell,y')\right|
\le B_\ell .
\label{eq:B}
\end{equation}
Here $y'$ specifies the other selected spins, and $B_\ell$ bounds
half the diagonal energy splitting between the two orientations of spin
$\ell$, while $h_\ell$ mixes its two orientations. Their competition
limits the record that this spin can hold, even when the other spins are
correlated. The host size, norm, geometry, and many-body gap are unrestricted.

The initial fringe visibility is the ground-state overlap. To express the
best recovery, also form the reduced states on the uncontrolled region:
\begin{align}
\Fglobal&=|\langle\Omega_-|\Omega_+\rangle|,\nonumber\\
\Fretain&=\left\|\sqrt{\rho_{R,-}}\sqrt{\rho_{R,+}}\right\|_1,
\quad\rho_{R,z}=\tr_M|\Omega_z\rangle\langle\Omega_z|.
\label{eq:fids}
\end{align}
We use root fidelity, so these quantities are amplitudes, not their
squares; $\|\cdot\|_1$ denotes the sum of singular values.
Uhlmann's theorem identifies $\Fretain$ with the largest overlap obtainable
by a unitary on the selected spins:
\begin{equation}
\Fretain=\max_{U_M}
\left|\langle\Omega_-|I_R\otimes U_M|\Omega_+\rangle\right|.
\label{eq:uhl}
\end{equation}
For the coherently prepared state
$(|0\rangle|\Omega_-\rangle+|1\rangle|\Omega_+\rangle)/\sqrt2$,
apply $U_M$ only on probe branch $|1\rangle$. The resulting fringe visibility
is the overlap in Eq.~\eqref{eq:uhl}. Thus $\Fglobal$ and $\Fretain$ are
respectively the initial and optimally recovered visibilities. The reduced
states are a way to compute this optimum; the protocol need not discard
or measure the environment.

To bound the optimum without finding $U_M$, we compare it with the overlap
of the two basis-outcome distributions on $R$:
\begin{equation}
F_Z=\sum_x\sqrt{p_-(x)p_+(x)},\qquad
p_z(x)=\sum_y\Omega_z(x,y)^2.
\label{eq:classical}
\end{equation}
This auxiliary quantity is the classical fidelity, or affinity, of those
distributions. The local Hamiltonian bounds all three overlaps:
\begin{gather}
\boxed{\eta_M F_Z\le\Fglobal\le\Fretain\le F_Z,}\nonumber\\
\eta_M=\prod_{\ell\in M}\frac{h_\ell}{\sqrt{h_\ell^2+B_\ell^2}}.
\label{eq:main}
\end{gather}
The last two inequalities follow from discarding $M$ and then measuring
$R$. The first is the additional constraint: even $F_Z$ can exceed the
initial overlap by at most $1/\eta_M$. In particular, the logarithmic gain
from optimal recovery obeys
\begin{equation}
\begin{gathered}
0\le\log\frac{\Fretain}{\Fglobal}\le\Lambda_M,
\\
\Lambda_M=\frac12\sum_{\ell\in M}\log\!\left[1+(B_\ell/h_\ell)^2\right].
\end{gathered}
\label{eq:log}
\end{equation}
For a fixed number of spins with uniformly bounded bias-to-field ratios,
$\Lambda_M$ remains uniformly bounded as the system grows. It limits every unitary on $M$,
including one tailored to the two states at each size. Multiplication by
a bounded factor cannot alter an algebraic exponent or an exponential
decay rate. The constant is sharp: an identical spectator host and
independent record spins $-h_\ell X_\ell\pm B_\ell Z_\ell$ have
$\Fretain=F_Z=1$ and $\Fglobal=\eta_M$.

\emph{Bounding conditional records.---}
The key step is to limit how different the two orientations of a selected
spin can be in either ground state, with all other coordinates held fixed.
For one spin, the two orientation sectors have Hamiltonians $C-f$ and
$C+f$, coupled by the transverse term $-hI$. Here $C$ describes the same
remaining system in both sectors, while the diagonal bias satisfies
$|f|\le B$. Starting from equal positive components, imaginary-time evolution
mixes the two sectors and keeps their component ratios between $r^{-1}$
and $r$, where $r=\exp[\operatorname{arsinh}(B/h)]$.
These bounds are preserved because
$2rB+h(1-r^2)=0$. Taking the ground-state limit gives
\begin{equation}
e^{-a_\ell}\le
\frac{\Omega_z(x,+_\ell,y')}{\Omega_z(x,-_\ell,y')}
\le e^{a_\ell},\qquad a_\ell=\operatorname{arsinh}(B_\ell/h_\ell).
\label{eq:cone}
\end{equation}

For each fixed $x$, write the normalized conditional amplitude on $M$ as
$v_{z,x}(y)=\Omega_z(x,y)/\sqrt{p_z(x)}$. Its squared components form a
probability distribution. Squaring Eq.~\eqref{eq:cone} bounds the odds of
the two orientations of spin $\ell$ by $e^{\pm2a_\ell}$; summing over
unobserved spins preserves that bound. The classical fidelity between
the two resulting binary distributions is at least
$\operatorname{sech}a_\ell$. Conditioning one spin at a time therefore gives
$\langle v_{-,x}|v_{+,x}\rangle\ge\prod_\ell\operatorname{sech}a_\ell
=\eta_M$, including when the selected spins are correlated.
Finally, positivity makes the full overlap a sum of these conditional
contributions without phase cancellations:
\begin{equation}
\Fglobal=\sum_x\sqrt{p_-(x)p_+(x)}\,
\langle v_{-,x}|v_{+,x}\rangle\ge\eta_M F_Z.
\label{eq:conditional-sum}
\end{equation}
The argument controls local component ratios rather than the response to
a perturbation of the entire ground state; no inverse many-body gap enters.
The Supplemental Material gives the imaginary-time and conditional-probability
steps in full.

\emph{Local recovery at criticality.---}
Consider a connected open Ising chain of even length $N=2L$ at its bulk
critical transverse field, with one bond whose sign labels the branches.
Reversing this bond changes the preferred relative orientation of its
neighboring spins while preserving the bulk critical point:
\begin{equation}
H_{s,N}=-\sum_{j=1}^{N}X_j
-\sum_{\substack{j=1\\j\ne L}}^{N-1}Z_jZ_{j+1}
-sZ_LZ_{L+1},\qquad s=\pm1.
\label{eq:ising}
\end{equation}
The half-chain flip $S_L=\prod_{j=1}^L X_j$ maps $H_{+,N}$ to
$H_{-,N}$. Both branches are therefore exactly critical and isospectral,
with lowest gap $\pi/N+O(N^{-2})$. Their positive ground states satisfy
$\Omega_{-,N}=S_L\Omega_{+,N}$, so the initial overlap $F_N=\Fglobal$ is the expectation of a spin-flip
string, conventionally called a disorder string:
\begin{equation}
F_N=\langle S_L\rangle_{+,N}
=A_*N^{-1/8}[1+O(N^{-1})],\qquad A_*>0.
\label{eq:criticalF}
\end{equation}
The exponent agrees with the Ising disorder-field dimension
\cite{Watts2001}. Here an exact finite-lattice covariance determinant,
its scalar product formula, and a uniform finite-strip estimate establish
Eq.~\eqref{eq:criticalF} without a scaling fit. Critical overlap exponents
also occur in other geometries, including bipartite fidelities
\cite{Dubail2011,Stephan2013}; the operation studied here is erasure of
selected physical spins in a chain whose bonds remain present.

The negative bond is diagonal and leaves all transverse spin flips equal
to $-1$, so positivity and Eq.~\eqref{eq:B} hold with $h=1$ and $B\le2$.
For \emph{any} set of $k$ physical spins, Eq.~\eqref{eq:main} gives
\begin{equation}
F_N\le F_R\le 5^{k/2}F_N.
\label{eq:critbound}
\end{equation}
For fixed $k$, even the optimized amplitude is $\Theta(N^{-1/8})$;
the optimized transition weight $Z_R=F_R^2$ is $\Theta(N^{-1/4})$.
The chosen sites need not be adjacent or fixed as $N$ changes.

\begin{figure}[t]
\includegraphics[width=\columnwidth]{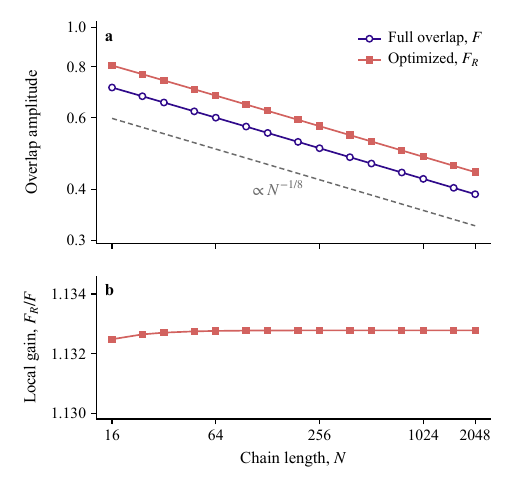}
\caption{Optimal local recovery preserves a critical orthogonality exponent.
For the connected sign-defect chain, the erased spins are the two adjacent
to the central bond, $M=\{L,L+1\}$.
(a) Bare overlap $F_N$ and optimal amplitude $F_R$, computed from the
physical two-spin transition matrix. The dashed reference has exponent
$1/8$, established analytically in Eq.~\eqref{eq:criticalF}.
(b) Their ratio approaches a finite enhancement in the displayed data.
All 15 sizes from $N=16$ to 2048 are shown; at the largest size
$F_R/F_N=1.13278$. Equation~\eqref{eq:critbound} bounds the ratio by
5 for every size, independently of this numerical trend.}
\label{fig:critical}
\end{figure}

For Fig.~\ref{fig:critical}, the selected region contains the two spins
touching the defect. We reconstruct
$C_M=\tr_R|\Omega_+\rangle\langle\Omega_-|$, the matrix whose contraction
$\tr(U_M C_M)$ is the recovery amplitude. Its trace norm is $F_R$ by
Eq.~\eqref{eq:uhl}. Reconstructing this physical-spin matrix allows every
two-spin unitary, including when $R$ consists of two disconnected intervals.
The enhancement is approximately $13.3\%$ at $N=2048$. The exact lattice
asymptotics and the recovery bound establish the common exponent; the
numerical optimization shows the finite enhancement actually achieved.

The bound also quantifies the resources needed to defeat the catastrophe.
For $\Fglobal=A N^{-\alpha}(1+o(1))$ and selected spins with
$B_\ell/h_\ell\le b$, $b>0$, achieving a fixed $F_R\ge f_0>0$ requires
\begin{equation}
|M|\ge\frac{2\alpha}{\log(1+b^2)}\log N+O(1).
\label{eq:cost}
\end{equation}
In the critical chain this is $|M|\ge\log N/(4\log5)+O(1)$.
More generally, $|M|=o(\log N)$ preserves the logarithmic exponent $1/8$,
although its prefactor need not remain bounded.
Equation~\eqref{eq:cost} is a necessary lower bound on the control cost.
A half-chain flip achieves full recovery at a linear cost.

\emph{Large recovery from auxiliary-spin control.---}
Local recovery can still produce a large enhancement at fixed parameters. Consider a core
chain with Hamiltonian $-K\sum_i Z_{A_i}Z_{A_{i+1}}-a\sum_iX_{A_i}$.
Attach two auxiliary spins, or leaves, $b_i,c_i$ to each core spin $A_i$.
The probe couples to two cells with opposite signs: set $s_0=1,s_1=-1$
and $s_i=0$ elsewhere. The leaf terms at cell $i$ are
\begin{equation}
-h(X_{b_i}+X_{c_i})
-[j_+Z_{A_i}+zg s_i]Z_{b_i}
-[j_-Z_{A_i}-zg s_i]Z_{c_i},
\label{eq:pendant}
\end{equation}
Here $j$ is the mean core--leaf coupling, $\epsilon$ its relative
mismatch between the two leaves, $h$ the leaf transverse field, and $g$
the probe-induced longitudinal field; $j_\pm=j(1\pm\epsilon)$.
The four driven leaves satisfy
Eq.~\eqref{eq:main} with $B_b=|j_+|+|g|$ and $B_c=|j_-|+|g|$.
This holds at finite $h$ and nonzero mismatch $\epsilon$.

\begin{figure}[t]
\includegraphics[width=\columnwidth]{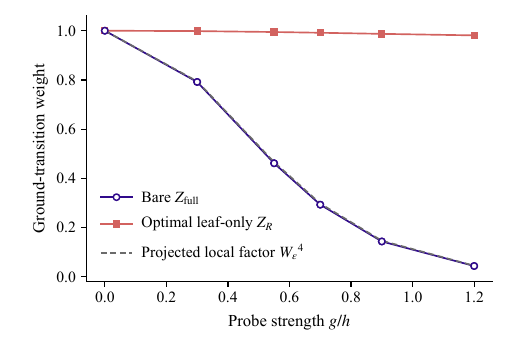}
\caption{A recoverable local record at finite mediator energy.
Complete ground states of a periodic four-cell (12-spin) pendant chain
give the bare transition weight $Z_{\rm full}=F_{\rm full}^2$ and optimal
leaf-only weight $Z_R=F_R^2$. Parameters are $K=1$, $h=40$, $j=20$,
$\epsilon=0.01$, and $a=1.25001500016$. At $g/h=1.2$,
$Z_{\rm full}=0.04379$ whereas $Z_R=0.98099$.
The dashed curve, $W_\epsilon^4$, is the squared overlap contributed by
the four driven leaves in their conditional ground states, derived in the
Supplemental Material. All displayed
weights refer to this finite system.}
\label{fig:local}
\end{figure}

Figure~\ref{fig:local} shows that optimal control of the four leaves raises
the squared overlap from approximately $0.044$ to $0.981$ in a complete
12-spin calculation, corresponding to fringe visibilities of $0.209$ and
$0.990$. The optimal recovery quantifies how much of this suppression
can be removed by access to the four leaves.

A concrete recovery operation can also be constructed. For each orientation
of a driven core spin, rotate its attached leaves from their branch-dependent
local ground states to the same reference spin state. Within this
conditional ground-state subspace, the rotations remove the leaves' branch
record. The full ground states also contain components outside that
subspace. When the minimum leaf excitation energy exceeds the total
driven-core flip scale, the Supplemental Material bounds the recovery
error using these energies and the core-flip matrix elements. The construction uses a local compression of the true state; its error
bound requires no gap in the remaining system. Local block reduction,
leakage control, and stability provide related tools
\cite{Bravyi2011,Marvian2015,Henheik2022}. Implementing the
core-conditioned rotations requires access to the driven cores as well as
their leaves. Equation~\eqref{eq:log} bounds the gain on that entire finite
support whenever the environment is collectively orthogonal.

\emph{Interferometric test and implications.---}
The critical-chain prediction can be tested by preparing its ground state
and an equal-weight probe superposition,
applying a probe-controlled half-chain flip to create the two branches,
and comparing probe fringes before and after recovery on $k$ chosen spins.
The half-chain operation prepares the branches; only the subsequent
recovery is restricted to $k$ spins. Repeating this protocol at different
sizes with fixed $k$ tests the predicted persistence of the exponent.
Impurity Ramsey experiments have measured dynamical many-body overlaps
\cite{Cetina2016}; single-qubit interferometry \cite{Dorner2013} and
individually addressed ion processors \cite{Schindler2013} provide relevant
readout and control methods. The static ground-state protocol proposed here
is derived in the Supplemental Material, including a reconstruction of the
optimal visibility from the $k$-spin transition matrix.

The auxiliary-spin example demonstrates that a weak finite-size signal
can contain a large recoverable contribution. The critical chain shows
a different limit: its orthogonality exponent survives even optimal local
recovery. More generally, within the stated positive-ground-state class
and uniformly bounded local ratios, an existing exponent constrains the
best control on any fixed number of spins. Maintaining nonvanishing
visibility then requires access to more spins, rather than only a better
operation on the same finite set. This control requirement follows from
local energy scales even when the many-body environment is gapless.

\emph{Acknowledgments}---This work is supported by Natural Science Foundation of Hebei Province (Grant No. A2026203033), Beijing National Laboratory for Condensed Matter Physics (Grant No. 2024BNLCMPKF020).

\bibliographystyle{apsrev4-2}
\bibliography{references}
\end{document}


\maketitle
\begin{abstract}
We prove a local amplitude-ratio bound for positive ground states and an
explicit reverse comparison for fidelity under finite-region erasure.
An exact lattice calculation for a connected critical Ising chain gives
amplitude exponent $1/8$, preserved by arbitrary optimal control on any
fixed spin set and, at the level of logarithmic exponents, on $o(\log N)$
spins. We derive the
finite Cauchy product and its uniform strip asymptotics, and document exact
two-spin optimization and an interferometric protocol for measuring the
recovery. A separate finite-mediator theorem quantifies local
compression and a decoder while retaining the rest of the environment
exactly. We also give a ground-preparation Ramsey bound, counterexamples
to stronger projected-state inferences, and independent full-Hamiltonian
calculations.
\end{abstract}

\section{Hamiltonian class and conventions}
Let $\mathcal H=\mathcal H_R\otimes(\mathbb C^2)^{\otimes k}$ be finite
dimensional, with product basis $(x,y)$ and $y_\ell=\pm1$. For $z=\pm$, take
\begin{equation}
H_z=H_{R,z}\otimes I_M-\sum_{\ell=1}^k h_\ell X_\ell+V_z(x,y),
\qquad h_\ell>0.
\label{eq:generalH}
\end{equation}
$H_{R,z}$ is real symmetric and has nonpositive off-diagonal matrix entries;
$V_z$ is a real diagonal matrix. Both full matrices are irreducible in the
\emph{same} basis. Perron--Frobenius therefore gives unique normalized ground
vectors $\Omega_z(x,y)>0$. There is no restriction on the norm or gap of
$H_R$, on a geometry in $R$, or on diagonal frustration. Diagonal interactions
among selected spins are allowed. Terms that flip selected spins jointly with
other spins, or make their transverse transition amplitudes configuration
dependent, are not part of Eq.~\eqref{eq:generalH}.

For every selected spin assume
\begin{equation}
\tfrac12|V_z(x,+_\ell,y')-V_z(x,-_\ell,y')|\le B_\ell,
\qquad B_\ell\ge0.
\label{eq:osc}
\end{equation}
The rest of $V_z$ can have extensive norm. In an Ising system, one may take
$B_\ell$ as the absolute longitudinal field plus the sum of absolute diagonal
couplings incident on that spin. Only bounds incident on the selected region
enter the theorem. Positive transverse fields may differ between branches:
then use $a_\ell=\max_z\operatorname{arsinh}(B_{\ell,z}/h_{\ell,z})$ below.
The main text uses common $h_\ell,B_\ell$ for a simpler expression.

All fidelities here are root fidelities. Define
\begin{align}
\Fglobal&=\langle\Omega_-|\Omega_+\rangle,\qquad
\rho_{R,z}=\tr_M|\Omega_z\rangle\langle\Omega_z|,\nonumber\\
\Fretain&=\tr\sqrt{\sqrt{\rho_{R,-}}\rho_{R,+}\sqrt{\rho_{R,-}}},\nonumber\\
p_z(x)&=\sum_y\Omega_z(x,y)^2,\qquad
F_Z=\sum_x\sqrt{p_-(x)p_+(x)}.
\label{eq:definitions}
\end{align}
Thus $Z=F^2$ denotes a squared fidelity or a ground-transition spectral
weight, never the amplitude $F$ itself.

\section{A local ground-amplitude bound}
The proof first compares the two orientations of one selected spin while
leaving every other degree of freedom in the Hamiltonian. The rest of the
system acts identically in the two orientation sectors; only the bounded
local bias distinguishes them. We show that imaginary-time evolution from
equal positive components preserves a finite ratio between the sectors,
which the ground state then inherits.
\begin{lemma}[A positive cone fixed by a local transverse field]
For every $\ell$, every branch, and every configuration of all other variables,
\begin{equation}
e^{-a_\ell}\le
\frac{\Omega_z(x,+_\ell,y')}{\Omega_z(x,-_\ell,y')}
\le e^{a_\ell},\qquad
 a_\ell=\operatorname{arsinh}(B_\ell/h_\ell).
\label{eq:ratio}
\end{equation}
\end{lemma}
\begin{proof}
Fix $\ell$ and suppress its subscript. Its two-block Hamiltonian is
\begin{equation}
H=\begin{pmatrix}C-f&-hI\\-hI&C+f\end{pmatrix},\qquad |f(x,y')|\le B.
\end{equation}
Here $C$ contains $H_R$, all other selected-spin flips, and the average of
the two diagonal potentials. It has nonpositive off-diagonal entries and is identical in both blocks. Set $u(t)=e^{-tH}\bm1=(u_+(t),u_-(t))$. Positivity of the semigroup
implies $u_\pm(t)>0$. For $r=\exp[\operatorname{arsinh}(B/h)]$, put
$d=u_+-ru_-$. Direct differentiation gives
\begin{equation}
\dot d=(-C+f-hrI)d+[2rf+h(1-r^2)I]u_-.
\label{eq:d}
\end{equation}
Because $2rB+h(1-r^2)=0$, the inhomogeneous term is componentwise
nonpositive. The homogeneous matrix has nonnegative off-diagonal entries,
so its exponential is entrywise nonnegative. Variation of constants and
$d(0)=(1-r)\bm1\le0$ show $d(t)\le0$. Reversing the roles of $+$ and $-$
gives $u_-(t)\le r u_+(t)$. Normalized imaginary-time evolution from the
strictly positive initial vector converges to the Perron ground vector.
Taking that finite-volume limit proves Eq.~\eqref{eq:ratio}.
\end{proof}
The argument does not estimate $E_0$, an extensive diagonal energy, a
many-body resolvent, or a spectral gap. In particular, it remains uniform
when host interactions or host matrix norms become arbitrarily large, provided
the selected coordinate oscillations in Eq.~\eqref{eq:osc} are bounded.

\section{Conditional affinity and the erasure theorem}
We now separate the information carried by the selected spins from the
amplitude distribution of the uncontrolled region. At each fixed $x$, the
normalized conditional vectors on $M$ describe the remaining difference
between the branches. The one-spin bound limits their distinguishability,
even when the selected spins are mutually correlated. For fixed $x$, write
\begin{equation}
\Omega_z(x,y)=\sqrt{p_z(x)}\,v_{z,x}(y),\qquad
\sum_y v_{z,x}(y)^2=1.
\end{equation}
The vectors $v_{z,x}$ are positive. Let $\pi_{z,x}(y)=v_{z,x}(y)^2$.
Their conditional odds at one spin lie between $e^{-2a_\ell}$ and
$e^{2a_\ell}$ by Eq.~\eqref{eq:ratio}. Summing the same pointwise
inequality over any subset of other spins preserves the odds bounds.
Consequently every sequential conditional probability for $y_\ell=+1$
lies in the interval
\begin{equation}
I_\ell=\left[\frac1{1+e^{2a_\ell}},
\frac{e^{2a_\ell}}{1+e^{2a_\ell}}\right].
\end{equation}
For two Bernoulli probabilities $p,q\in I_\ell$,
\begin{equation}
\sqrt{pq}+\sqrt{(1-p)(1-q)}\ge\sech a_\ell.
\label{eq:binary}
\end{equation}
Indeed this function is concave in each variable separately, so its minimum
over the square lies at a vertex; opposite endpoints give $\sech a_\ell$.

To apply Eq.~\eqref{eq:binary} to correlated selected spins, use the chain
rule for both $\pi_{z,x}$. At the last spin, sum the square root of the two
conditional probabilities for each fixed prefix; this is at least
$\sech a_k$. Repeat for the previous spin. Induction gives
\begin{equation}
\sum_y\sqrt{\pi_{-,x}(y)\pi_{+,x}(y)}
=\langle v_{-,x}|v_{+,x}\rangle
\ge\prod_{\ell=1}^k\sech a_\ell\equiv\eta_M.
\label{eq:cond}
\end{equation}
No independence of the selected spins has been assumed.

\begin{theorem}[Full finite-parameter fidelity comparison]
\label{thm:erasure}
For Eq.~\eqref{eq:generalH} under the stated assumptions,
\begin{equation}
\boxed{\eta_M F_Z\le\Fglobal\le\Fretain\le F_Z.}
\label{eq:theorem}
\end{equation}
In the common-parameter case,
\begin{equation}
\eta_M=\prod_{\ell=1}^k\frac{h_\ell}{\sqrt{h_\ell^2+B_\ell^2}}.
\end{equation}
\end{theorem}
\begin{proof}
Multiply Eq.~\eqref{eq:cond} by $\sqrt{p_-(x)p_+(x)}$ and sum over $x$.
This proves the first inequality. The last two follow from root-fidelity
monotonicity under partial trace and measurement, respectively
\cite{Uhlmann1976,Jozsa1994}. Equivalently, Uhlmann's variational formula
gives $\Fretain$ as the largest overlap obtainable by a unitary on $M$, so
it is at least $\Fglobal$. Measuring $R$ gives the classical affinity $F_Z$.
\end{proof}

\subsection{Sharpness and counterexamples}
Take the same pure spectator state on $R$, and independent selected-spin
Hamiltonians $-h_\ell X_\ell\pm B_\ell Z_\ell$. Their branch spinors have
overlap $h_\ell/\sqrt{h_\ell^2+B_\ell^2}$, so $\Fretain=F_Z=1$ and
$\Fglobal=\eta_M$. This saturates the lower constant in the full model class.
It does not show saturation for fixed pendant parameters.

The positivity hypothesis cannot be dropped. The two Bell states
$(|00\rangle\pm|11\rangle)/\sqrt2$ are orthogonal, while tracing either
spin gives the same maximally mixed state. Nor is finite dimension alone
sufficient for a uniform constant: one selected spin with $h/B\to0$ has
vanishing overlap although its retained spectator is unchanged. Finally,
$k$ independent local records give $\Fglobal=\eta^k$ with $\Fretain=1$;
if $k$ grows, erasure can indeed remove an extensive fidelity loss.

\section{Orthogonality exponents and optimal local readout}
Theorem~\ref{thm:erasure} implies
\begin{equation}
0\le\log\frac{\Fretain}{\Fglobal}\le
\Lambda_M=-\log\eta_M
=\frac12\sum_\ell\log[1+(B_\ell/h_\ell)^2].
\label{eq:cost}
\end{equation}
The same logarithmic bound holds with $F_Z$ in place of $\Fretain$.
For fixed $k$ and uniformly bounded bias-to-field ratios, the three
fidelities vanish together or have a strictly positive lower limit together.
If $a_N\to\infty$ and $\Lambda_{M_N}=o(a_N)$, every existing limit of
$-\log F/a_N$ is the same for all three. Taking $a_N=\log N$ gives equality
of algebraic amplitude exponents; taking $a_N=N$ gives equality of exponential
rates. This is a comparison of finite-volume sequences. It does not assume
or establish an infinite-volume real-time dynamics.

Uhlmann's theorem gives the operational identity
\begin{equation}
\Fretain=\max_{U_M}\left|\langle\Omega_-|I_R\otimes U_M|\Omega_+\rangle\right|.
\label{eq:uhlmann}
\end{equation}
The optimization allows a different, state-dependent unitary at each size.
Even that freedom cannot produce a nonzero ground-transition weight from a
catastrophe by acting on a fixed region obeying the uniform local bounds.
If $\Fglobal=A N^{-\alpha}(1+o(1))$, $\Fretain\ge f_0>0$, and
$B_\ell/h_\ell\le b$ with $b>0$, then
\begin{equation}
|M_N|\ge\frac{2\alpha}{\log(1+b^2)}\log N+O(1).
\end{equation}
A rate $\Fglobal\sim e^{-sN}$ similarly requires a linear number of
controlled spins. These are necessary bounds, not constructive recovery
protocols. Known orthogonality theory concerns when a particular model has a
nonzero exponent \cite{Anderson1967,Munder2012}; Eq.~\eqref{eq:cost} instead
says which part of that exponent can be removed by local readout.

\section{A connected critical chain with an exact exponent}
\label{sec:critical-example}

Consider the open chain of $N=2L$ spins,
\begin{equation}
H_{s,N}=-\sum_{j=1}^{N}X_j
-\sum_{\substack{j=1\\j\ne L}}^{N-1}Z_jZ_{j+1}
-sZ_LZ_{L+1},\qquad s=\pm1.
\label{eq:crit-H}
\end{equation}
All transverse fields and bond magnitudes equal one. Both chains are
connected; the minus branch has one antiferromagnetic bond. In the product
$Z$ basis, the off-diagonal entries are the single-spin flips $-1$, so each
finite Hamiltonian has a unique normalized ground vector
$\Omega_{s,N}>0$. Define $F_N=\langle\Omega_{-,N}|\Omega_{+,N}\rangle$.
For a spin set $M_N$ and its complement $R_N$, write $F_{R,N}$ for the root
fidelity after tracing $M_N$.

\begin{theorem}[Critical orthogonality under optimal local control]
\label{thm:critical}
There is a constant $A_*>0$ such that
\begin{equation}
F_N=A_*N^{-1/8}[1+O(N^{-1})].
\label{eq:crit-exponent}
\end{equation}
For every $M_N$ with $|M_N|=k_N$,
\begin{equation}
F_N\le F_{R,N}\le5^{k_N/2}F_N,
\qquad
F_{R,N}=\max_{U_{M_N}}
|\langle\Omega_{-,N}|U_{M_N}|\Omega_{+,N}\rangle|.
\label{eq:crit-erasure}
\end{equation}
Hence fixed $k_N$ gives $F_{R,N}=\Theta(N^{-1/8})$ and an optimal squared
transition weight $\Theta(N^{-1/4})$. More generally,
\begin{equation}
k_N=o(\log N)\quad\Longrightarrow\quad
\lim_{N\to\infty}\frac{-\log F_{R,N}}{\log N}=\frac18.
\label{eq:crit-sublog}
\end{equation}
The spin set and optimizing unitary may depend on $N$ and on both states.
\end{theorem}

\subsection{Gauge transformation and the disorder string}
Put $S_m=\prod_{j=1}^m X_j$. Conjugation by $S_L$ reverses the central
bond and leaves every other term unchanged:
\begin{equation}
H_{-,N}=S_LH_{+,N}S_L,\qquad
\Omega_{-,N}=S_L\Omega_{+,N},\qquad
F_N=\langle S_L\rangle_{+,N}>0.
\label{eq:crit-string}
\end{equation}
The ground-vector identity follows from uniqueness and positivity because
$S_L$ is a permutation matrix in the $Z$ basis. It needs no choice of a
symmetry-broken thermodynamic state. The branches are isospectral to the
uniform critical Ising chain. Their one-fermion excitation energies are
\begin{equation}
\varepsilon_q=4\sin\frac{(2q-1)\pi}{4N+2},\qquad q=1,\ldots,N,
\label{eq:crit-spectrum}
\end{equation}
so their gap is $\pi/N+O(N^{-2})$. The string in
Eq.~\eqref{eq:crit-string} is a disorder field ending at an open boundary
\cite{Watts2001}. We derive its exponent directly on the lattice below.

\subsection{Finite covariance and the Cauchy product}
The half-chain spin flip expresses the branch overlap as a string
expectation in one ground state. To evaluate that expectation exactly,
we use the free-fermion representation of the uniform Ising chain.
This changes the calculation, while the region available for recovery
remains a set of physical spins. Use the Jordan--Wigner Majoranas
\begin{equation}
a_{2j-1}=\left(\prod_{r<j}X_r\right)Z_j,\qquad
a_{2j}=\left(\prod_{r<j}X_r\right)Y_j.
\label{eq:crit-majoranas}
\end{equation}
They obey $ia_{2j-1}a_{2j}=X_j$,
$ia_{2j}a_{2j+1}=Z_jZ_{j+1}$, and
$H_{+,N}=-i\sum_{r=1}^{2N-1}a_ra_{r+1}$. Set
$\vartheta=\pi/(2N+1)$. The mixed covariance is
\begin{align}
G_{jk}&=\langle ia_{2k-1}a_{2j}\rangle_{+,N}\nonumber\\
&=\frac{4}{2N+1}\sum_{q=1}^N
\sin[(2q-1)\vartheta j]\cos[(2q-1)\vartheta(k-\tfrac12)]\nonumber\\
&=\frac1{2N+1}\left\{
\csc[(j-k+\tfrac12)\vartheta]+
\csc[(j+k-\tfrac12)\vartheta]\right\}.
\label{eq:crit-covariance}
\end{align}
The sine and cosine functions are the singular vectors of the critical
bidiagonal hopping matrix, with singular values
$2\sin[(2q-1)\vartheta/2]$. Summing the finite sine series gives the last
line. Since $S_m=\prod_{j=1}^m(ia_{2j-1}a_{2j})$, its expectation is a
$2m$-Majorana correlation. Contractions between distinct Majoranas of the
same index parity vanish. Wick's Pfaffian therefore reduces to the
determinant of the mixed covariance block:
\begin{equation}
D_{m,N}:=\langle S_m\rangle_{+,N}
=\det(G_{jk})_{1\le j,k\le m},\qquad D_{0,N}=1.
\label{eq:crit-determinant}
\end{equation}

The identity
$\csc(a-b)+\csc(a+b)=2\sin a\cos b/(\sin^2a-\sin^2b)$
puts this determinant in Cauchy form. With
$x_j=\sin^2(j\vartheta)$ and
$y_k=\sin^2[(k-\tfrac12)\vartheta]$, the positive finite product for
$m\le N/2$ is
\begin{equation}
D_{m,N}=\left(\frac2{2N+1}\right)^m
\left[\prod_{j=1}^{m}\sin(j\vartheta)
\cos((j-\tfrac12)\vartheta)\right]
\frac{\displaystyle\prod_{1\le j<\ell\le m}(x_\ell-x_j)
\prod_{1\le k<\ell\le m}(y_\ell-y_k)}
{\displaystyle\prod_{j,k=1}^{m}|x_j-y_k|}.
\label{eq:crit-cauchy-product}
\end{equation}
Here $y_1<x_1<y_2<x_2<\cdots<y_m<x_m$; the absolute values collect
the compensating signs in the Cauchy determinant. Dividing consecutive
products and using
$\sin^2a-\sin^2b=\sin(a+b)\sin(a-b)$ gives
\begin{equation}
\boxed{R_{m,N}:=\frac{D_{m,N}}{D_{m-1,N}}
=\frac1{(2N+1)\sin[(2m-\tfrac12)\vartheta]}
\prod_{r=1}^{2m-1}
\left\{\frac{\sin(r\vartheta)}
{\sin[(r-\tfrac12)\vartheta]}\right\}^{\!2}.}
\label{eq:crit-recurrence}
\end{equation}
All factors in this recurrence are positive in the required range. Cached
cumulative logarithms of its sine ratios give an $O(N)$ evaluation of
$D_{N/2,N}$.

\subsection{Half-line asymptotics and the uniform finite-strip correction}
For fixed $m$, taking $N\to\infty$ in
Eq.~\eqref{eq:crit-recurrence} gives
\begin{equation}
R_m^\infty=\frac{D_{m,\infty}}{D_{m-1,\infty}}
=\frac{\Gamma(2m)^2}
{\Gamma(2m+\tfrac12)\Gamma(2m-\tfrac12)}.
\label{eq:crit-gamma-ratio}
\end{equation}
The gamma expansion implies
\begin{equation}
\log R_m^\infty=-\frac1{8m}+O(m^{-2}),\qquad
D_{m,\infty}=A m^{-1/8}[1+O(m^{-1})],
\label{eq:crit-halfline}
\end{equation}
where a convergent definition of the positive constant is
\begin{equation}
\log A=-\frac{\gamma_E}{8}
+\sum_{m=1}^{\infty}\left[
2\log\Gamma(2m)-\log\Gamma(2m+\tfrac12)
-\log\Gamma(2m-\tfrac12)+\frac1{8m}\right].
\label{eq:crit-constant}
\end{equation}
The summand is $O(m^{-2})$, so $0<A<\infty$. To apply this result at
$m=N/2$, we next control the correction uniformly in the string length.

Let $f(t)=\log(\sin t/t)$ with $f(0)=0$. Dividing the finite recurrence
by Eq.~\eqref{eq:crit-gamma-ratio} gives the exact identity
\begin{equation}
\log\frac{R_{m,N}}{R_m^\infty}
=2\sum_{r=1}^{2m-1}
\{f(r\vartheta)-f[(r-\tfrac12)\vartheta]\}
-f[(2m-\tfrac12)\vartheta].
\label{eq:crit-strip-ratio}
\end{equation}
The derivatives of $f$ are bounded on $[0,\pi/2]$ and $f'(0)=0$.
A shifted Euler--Maclaurin estimate, uniformly for $n\vartheta\le\pi/2$,
is
\begin{equation}
\sum_{r=1}^{n}\{f(r\vartheta)-f[(r-\tfrac12)\vartheta]\}
=\frac12f(n\vartheta)+\frac{\vartheta}{8}f'(n\vartheta)
+O(\vartheta^2).
\label{eq:crit-euler-maclaurin}
\end{equation}
For clarity, Taylor expansion over each half-interval gives
$(\vartheta/2)\sum_r f'(r\vartheta)
-(\vartheta^2/8)\sum_r f''(r\vartheta)+O(n\vartheta^3)$.
The trapezoid estimates for these two sums give
Eq.~\eqref{eq:crit-euler-maclaurin}; bounded derivatives and
$n\vartheta\le\pi/2$ make the remainder uniform.

Substitute $n=2m-1$ into Eq.~\eqref{eq:crit-strip-ratio} and expand the
last half-step. This yields
\begin{equation}
\log\frac{R_{m,N}}{R_m^\infty}
=-\frac{\vartheta}{4}f'(2m\vartheta)+O(\vartheta^2).
\label{eq:crit-strip-step}
\end{equation}
Summing from $1$ to $m$ and using a Riemann-sum estimate with step
$2\vartheta$ proves
\begin{equation}
\log\frac{D_{m,N}}{D_{m,\infty}}
=-\frac18 f(2m\vartheta)+O(N^{-1}),\qquad m\le N/2,
\label{eq:crit-strip-uniform}
\end{equation}
uniformly over that range. Combining this with
Eq.~\eqref{eq:crit-halfline}, for $m/N\to x\in(0,1/2]$ we obtain
\begin{equation}
D_{m,N}=A\left[
\frac{2N+1}{2\pi}\sin\frac{2\pi m}{2N+1}
\right]^{-1/8}
[1+O(m^{-1})+O(N^{-1})].
\label{eq:crit-strip-asymptotic}
\end{equation}
At $m=N/2$, the sine equals $1+O(N^{-2})$. Thus
Eq.~\eqref{eq:crit-string} gives Eq.~\eqref{eq:crit-exponent} with
$A_*=A\pi^{1/8}$. This finite-strip estimate justifies the simultaneous
large-$m$, large-$N$ limit.

\subsection{Erasure bound and the necessary control region}
For either branch of Eq.~\eqref{eq:crit-H}, the diagonal half oscillation
at a spin is bounded by the sum of its incident bond magnitudes. Hence
$h_j=1$, $B_j=2$ in the interior, and $B_j=1$ at an endpoint. Applying
Theorem~\ref{thm:erasure} gives
\begin{equation}
\eta_{M_N}=\prod_{j\in M_N}(1+B_j^2)^{-1/2}
\ge5^{-k_N/2},\qquad
\eta_{M_N}F_{R,N}\le F_N\le F_{R,N}.
\label{eq:crit-local-budget}
\end{equation}
Together with Uhlmann's identity, this proves
Eq.~\eqref{eq:crit-erasure}. For the central pair $M_N=\{L,L+1\}$,
the upper multiplier is $5$. Taking logarithms gives
Eq.~\eqref{eq:crit-sublog} and completes the proof of
Theorem~\ref{thm:critical}. A particular local matrix element can vanish
by symmetry; the theorem concerns the maximum over all local unitaries.

A fixed target $F_{R,N}\ge f_0>0$ requires
\begin{equation}
k_N\ge\frac{\log N}{4\log5}+O(1).
\label{eq:crit-control-cost}
\end{equation}
For $k_N\le c\log N$, the same inequality yields
$F_{R,N}=O(N^{-1/8+c\log5/2})$. Equation~\eqref{eq:crit-control-cost}
is a necessary count. An explicit sufficient operation is the half-chain
string: taking $M_N=\{1,\ldots,L\}$ and $U_{M_N}=S_L$ gives amplitude
one. These results do not identify the smallest sufficient region.

The sign-defect geometry differs from cutting and rejoining the central
bond. For the open critical Ising chain with central bond $0$ versus $1$,
the bipartite-fidelity result gives
$-\log F^2=(c/8)\log N+O(1)$ with $c=1/2$, hence amplitude exponent
$1/32$ \cite{Dubail2011,Stephan2013}. The known disorder-field and Cauchy
methods provide the full-state exponent used here; the erasure comparison
fixes the exponent after arbitrary optimal control on finitely many spins.

\section{Exact numerical optimization on the central two spins}
\label{sec:critical-numerics}

We evaluated Eq.~\eqref{eq:crit-H} by free-fermion linear algebra and
reconstructed the transition operator on the two physical spins
$M=\{L,L+1\}$. This gives the optimal spin fidelity even though the retained
region consists of two disjoint intervals.

\subsection{Ground-state polar matrices and overlap}
Order the Majoranas as
$(a_1,a_3,\ldots,a_{2N-1};a_2,a_4,\ldots,a_{2N})$.
Writing $H=(i/4)\boldsymbol a^T\mathcal A\boldsymbol a$, use
\begin{equation}
\mathcal A=\begin{pmatrix}0&-2T\\2T^T&0\end{pmatrix},\qquad
T_{jj}=1,\quad T_{j+1,j}=-J_j,
\label{eq:crit-numeric-hopping}
\end{equation}
where $J_L=s$ and all other $J_j=1$. If $T=U\Sigma V^T$, let
$Q=UV^T$. The covariance, with
$\Gamma_{pq}=i\langle a_pa_q\rangle$ for distinct indices, is
\begin{equation}
\Gamma=\begin{pmatrix}0&Q\\-Q^T&0\end{pmatrix}.
\label{eq:crit-numeric-covariance}
\end{equation}
For the two branches $a=-$, $b=+$,
\begin{equation}
F_N=\left|\det\frac{Q_a+Q_b}{2}\right|^{1/2}.
\label{eq:crit-numeric-overlap}
\end{equation}
The implementation uses a logarithmic LU determinant. The gauge identity
is $Q_-=D Q_+D$, with $D=-1$ on the left half and $D=+1$ on the right.

\subsection{Spin transition matrix and optimal unitary}
Define the normalized transition operator on the erased spins,
\begin{equation}
\tau_M=\frac{\tr_R|\Omega_b\rangle\langle\Omega_a|}{F_N},\qquad
F_{R,N}=F_N\|\tau_M\|_1.
\label{eq:crit-transition}
\end{equation}
The second equality is Uhlmann optimization. In particular, if
$\tau_M=W\Sigma V^\dagger$ is a singular-value decomposition, then
$U_M=VW^\dagger$ attains
$|\tr(U_M\tau_M)|=\tr\Sigma$. Thus the computation allows every
unitary on these two spins.

The normalized transition contractions
$C_{pq}=\langle\Omega_a|a_pa_q|\Omega_b\rangle/F_N$ are
\begin{equation}
B=2(Q_a+Q_b)^{-T},\qquad
C=\begin{pmatrix}
B Q_b^T&-iB\\ iB^T&Q_a^T B
\end{pmatrix}.
\label{eq:crit-transition-contractions}
\end{equation}
They satisfy $C+C^T=2I$ and the vacuum constraints
$(I+i\Gamma_a)C=0$, $C(I+i\Gamma_b)=0$.
LU solves evaluate the rows and columns required for the two erased sites
without forming the full inverse.

Both ground states have even global $X$ parity. Every parity-odd Pauli
moment of $\tau_M$ therefore vanishes. Every parity-even two-spin Pauli
operator is the identity or a product of two or four local Majoranas;
the external Jordan--Wigner strings cancel. Wick Pfaffians of the local
$C$ submatrix give its transition moment. With $\sigma^0=I$, the expansion
\begin{equation}
\tau_M=\frac14\sum_{\mu,\nu=0,x,y,z}
\frac{\langle\Omega_a|\sigma_L^\mu\sigma_{L+1}^\nu
|\Omega_b\rangle}{F_N}
\,\sigma_L^\mu\sigma_{L+1}^\nu
\label{eq:crit-pauli-reconstruction}
\end{equation}
reconstructs the full $4\times4$ spin matrix. Its nonsymmetric part is
retained; $\tau_M$ need not be Hermitian or positive. Its singular values
give the norm in Eq.~\eqref{eq:crit-transition}. This construction uses
physical spin moments on $M$, rather than a fermionic covariance assigned
to the disconnected retained region.

\subsection{Independent checks and numerical results}
An independent sparse calculation built the Hamiltonian directly in the
product $Z$ basis for $N=4,6,8,10,12,14$. Reshaping each ground vector
into a matrix indexed by $M$ and $R$ gives
$\tr_R|\Omega_b\rangle\langle\Omega_a|$ by direct matrix multiplication.
For the sign-defect pair, the largest absolute differences from the
free-fermion computation were $4.99\times10^{-13}$ in $F_N$,
$2.73\times10^{-13}$ in $F_{R,N}$, and $3.84\times10^{-13}$ in any entry
of $\tau_M$. This compares the entire physical-spin transition operator,
including the Jordan--Wigner signs and the determinant square root.

The large-chain calculations used $N=16$, $24$, $32$, $48$, $64$, $96$,
$128$, $192$, $256$, $384$, $512$, $768$, $1024$, $1536$, and $2048$.
All cases obeyed $1\le F_{R,N}/F_N\le5$, $F_{R,N}\le1$, unit trace of
$\tau_M$, and the anticommutation check on $C$. At $N=2048$,
\begin{equation}
F_N=0.389604230344,\qquad
F_{R,N}=0.441336736855,\qquad
\frac{F_{R,N}}{F_N}=1.1327821992.
\label{eq:crit-numeric-endpoint}
\end{equation}
Fits to the last five sizes, $N\ge512$, gave exponent $0.1249349651$ for
$F_N$ and $0.1249347890$ for $F_{R,N}$. The direct doubling exponents
from $N=1024$ to $2048$ were $0.1249559356$ and $0.1249558616$.
These finite-size values illustrate the exponent proved in
Sec.~\ref{sec:critical-example} and the finite gain from optimal control.

The calculations used Python 3.14.3, NumPy 2.4.2, and SciPy 1.18.0.
The accompanying numerical data contain the model parameters, overlaps,
and full spin transition matrices for all 15 sizes; the analysis code
reproduces the covariance and physical-spin calculations described above.

\section{Interferometric meaning of local recovery}
\label{sec:interferometry}
The overlaps in the erasure theorem can be measured as probe fringe
visibilities. Prepare a two-state probe and its environment in
\begin{equation}
|\Psi\rangle=\frac{|0\rangle|\Omega_-\rangle+
|1\rangle|\Omega_+\rangle}{\sqrt2},
\end{equation}
and apply the controlled operation
\begin{equation}
C_U=|0\rangle\langle0|\otimes I+
|1\rangle\langle1|\otimes(I_R\otimes U_M).
\end{equation}
A phase scan of the final probe pulse gives
\begin{equation}
p_0(\phi)=\frac12\left[1+\operatorname{Re}(e^{i\phi}c_U)\right],
\qquad c_U=\langle\Omega_-|I_R\otimes U_M|\Omega_+\rangle.
\label{eq:probe-fringe}
\end{equation}
The visibility is therefore $\mathcal V(U_M)=|c_U|$. The usual relation
between branch distinguishability and interference \cite{Englert1996},
together with Uhlmann's theorem \cite{Uhlmann1976,Jozsa1994}, yields
\begin{equation}
\mathcal V(I)=\Fglobal,\qquad
\max_{U_M}\mathcal V(U_M)=\Fretain,\qquad
1\le\frac{\max_{U_M}\mathcal V(U_M)}{\mathcal V(I)}\le e^{\Lambda_M}.
\label{eq:probe-recovery}
\end{equation}
Quantum erasure and finite-access recovery have established formulations
\cite{Scully1982,Miatto2015}; the last inequality supplies the
Hamiltonian-dependent restriction studied here. A probe-independent,
trace-preserving operation on $M$, including discarding it, leaves the
probe's reduced state unchanged. The recovery in
Eq.~\eqref{eq:probe-recovery} instead uses probe-conditioned control.

\subsection{A finite-size test in the critical chain}
For the connected sign-defect chain,
$|\Omega_-\rangle=S_L|\Omega_+\rangle$ with
$S_L=\prod_{j=1}^L X_j$. Starting from $|\Omega_+\rangle$ and the probe
state $(|0\rangle+|1\rangle)/\sqrt2$, a half-chain flip conditioned on
probe state $0$ prepares $|\Psi\rangle$ exactly in the ideal gate model.
A subsequent $C_U$ and probe phase scan measure the recovery visibility.
For every fixed number $k$ of controlled spins, the theorem predicts
\begin{equation}
A_*N^{-1/8}[1+O(N^{-1})]\le\mathcal V_{\max}
\le5^{k/2}A_*N^{-1/8}[1+O(N^{-1})].
\end{equation}
The half-chain operation prepares the input branches; the spatial
restriction applies to the subsequent recovery. Preparation accuracy and
gate errors must be characterized at each size. The decreasing critical
gap can increase the preparation time.

This is a prospective measurement of static ground-state overlaps.
Impurity Ramsey experiments have measured time-dependent overlaps in a
Fermi sea \cite{Cetina2016}. Single-qubit interferometry
\cite{Dorner2013} and individually addressed trapped-ion processors
\cite{Schindler2013} provide readout and control methods relevant to the
proposed sequence; those works do not implement the present critical-chain
recovery test. Its size-dependent visibility differs from the time-dependent
amplitude $\ell_{\mathcal K}(t)$ considered later in this supplement.

\subsection{Reconstructing the optimal visibility on a small region}
Optimization requires only the transition operator on $M$,
\begin{equation}
C_M=\tr_R|\Omega_+\rangle\langle\Omega_-|,
\qquad
C_M=2^{-k}\sum_{P_M}\tr(P_M C_M)P_M,
\end{equation}
where the sum runs over all $4^k$ Pauli strings on the selected spins.
In the critical-chain numerical convention of
Eq.~\eqref{eq:crit-transition}, $C_M=F_N\tau_M$; the interferometric
reconstruction and the numerical optimization therefore use the same
operator, with different normalizations.
Each coefficient is the complex interference amplitude
$\tr(P_M C_M)=\langle\Omega_-|P_M|\Omega_+\rangle$.
Controlled Pauli operations followed by two probe quadrature measurements
determine its real and imaginary parts. This reconstructs a
$2^k\times2^k$ matrix rather than the full environment state. Its singular
values give $\Fretain=\|C_M\|_1$; if $C_M=W\Sigma V^\dagger$, then
$U_M=VW^\dagger$ attains the optimum.

For fixed $k$, the number of distinct Pauli amplitudes is independent of
$N$, although the precision and number of repetitions needed to resolve a
decreasing visibility can grow. In particular, a full trace-norm error
$\|\widetilde\rho-|\Psi\rangle\langle\Psi|\|_1\le d$ changes any
unit-norm probe quadrature, and hence the visibility, by at most $d$.
Vanishing relative systematic error in the critical size law is guaranteed
if $d=o(N^{-1/8})$. Finally, the visibility is the amplitude $F$; the
ground-transition spectral weight $Z=F^2$ shown for the auxiliary-spin
example is its square.

\section{Finite-mediator compression with an arbitrary conditional host}
\label{sec:compression}
The erasure theorem limits the best possible recovery. We next construct
a specific recovery operation for the auxiliary-spin model and estimate
its error. The construction aligns each leaf's conditional ground state
between the two probe branches. Its accuracy is set by how much of the
exact state lies outside those local ground-state subspaces.
The next statement does not require real positivity, but uses the independent
pendant architecture. Let there be $m$ driven core spins $A_i$ and two leaves
at each driven cell, with
\begin{align}
H_z={}&H_R(\{Z_{A_i}\})-\sum_i a_i X_{A_i}
-\sum_i h_i(X_{b_i}+X_{c_i})\nonumber\\
&-\sum_i\{(J_{b,i}Z_{A_i}+zg_i)Z_{b_i}
 +(J_{c,i}Z_{A_i}-zg_i)Z_{c_i}\}.
\label{eq:pendant}
\end{align}
Here $a_i\ge0$, $h_i>0$. $H_R$ commutes with all driven $Z_{A_i}$;
its conditional blocks on the retained host can be arbitrary Hermitian
matrices, including complex, mutually noncommuting blocks of arbitrary norm.
All undriven cells of an original decorated chain are part of $H_R$ and
remain exact. Only the $2m$ driven leaves will be compressed. For this
possibly complex Hamiltonian class, define
$\Fglobal=|\langle\Omega_-|\Omega_+\rangle|$; the absolute value makes
the fidelity independent of the ground-vector phases.

For a fixed longitudinal field $f$, the leaf Hamiltonian $-h_iX-fZ$
has ground energy $-E_i(f)$, with $E_i(f)=\sqrt{h_i^2+f^2}$, and
normalized positive ground spinor
\begin{equation}
|n_i(f)\rangle=
\begin{pmatrix}\sqrt{[E_i(f)+f]/[2E_i(f)]}\\
\sqrt{[E_i(f)-f]/[2E_i(f)]}\end{pmatrix}.
\end{equation}
For a fixed core orientation $\sigma_i=\pm1$, the two leaves experience
fields $J_{b,i}\sigma_i+zg_i$ and $J_{c,i}\sigma_i-zg_i$.
The isometry $\mathcal E_z$ embeds the retained system into the full spin
space by attaching the corresponding two spinors at every driven cell.
It preserves the core coordinates and acts identically on all other
retained degrees of freedom. Two local overlaps control this embedding:
$W_i$ compares the leaves in opposite probe branches at fixed core
orientation, while $R_i$ compares them before and after a core flip in
one branch. Explicitly,
\begin{align}
W_i&=\prod_{J=J_{b,i},J_{c,i}}
\sqrt{\frac12\left[1+\frac{h_i^2+J^2-g_i^2}{E_i(J+g_i)E_i(J-g_i)}\right]},\label{eq:W}\\
R_i&=\prod_{J=J_{b,i},J_{c,i}}
\sqrt{\frac12\left[1+\frac{h_i^2+g_i^2-J^2}{E_i(J+g_i)E_i(J-g_i)}\right]}.
\label{eq:R}
\end{align}
Taking the dot product of the corresponding Bloch vectors proves these
formulas. Positivity fixes the signs of the square roots. The first overlap
is independent of the core value and the drive sign, so
\begin{equation}
\mathcal E_-^\dagger \mathcal E_+=cI,\qquad c=\prod_i W_i.
\label{eq:scalar}
\end{equation}
For two identical driven cells, $c=W_\epsilon^2$ and projected squared
overlap is $W_\epsilon^4$ times the retained squared overlap. This scalar
identity is inherited from the preceding projected-overlap analysis; it is
not asserted to hold between arbitrary full finite-$h$ ground states.

The projector $P_z=\mathcal E_z\mathcal E_z^\dagger$ selects the subspace
in which all driven leaves occupy these conditional ground states.
Its complement $Q_z=I-P_z$ contains at least one leaf excitation.
The following local scales quantify the cost of leaving that subspace:

\begin{align}
\Delta_0&=2\min_{i,J,\varsigma=\pm1}E_i(J+\varsigma g_i),&
r&=\sum_i a_i,\nonumber\\
\delta&=\Delta_0-r,&
b_*^2&=\sum_i a_i^2(1-R_i^2).
\label{eq:constants}
\end{align}
Here $\Delta_0$ is the smallest conditional leaf excitation energy,
$r$ bounds the total driven-core flip term, and $b_*$ is the norm of the
coupling between $P_z$ and $Q_z$. As proved below, $\delta$ is a lower
bound on the energy separation of the discarded sector above the full
ground energy. Assume $\delta>0$.
For a normalized full ground vector $\Omega_z$ define
\begin{equation}
\psi_z=\frac{\mathcal E_z^\dagger\Omega_z}{\|\mathcal E_z^\dagger\Omega_z\|},\qquad
F_T=|\langle\psi_-|\psi_+\rangle|.
\end{equation}
These are compressed \emph{true} states, not projected-Hamiltonian ground
states. This distinction is essential near a gapless retained system.

\begin{theorem}[Gap-free local compression]
\label{thm:compression}
Let $\theta=\arctan(b_*/\delta)$. Then
\begin{align}
q_z=\|Q_z\Omega_z\|^2&\le\frac{b_*^2}{\delta^2+b_*^2}=\sin^2\theta,
\label{eq:leakage}\\
|\arccos\Fglobal-\arccos(cF_T)|&\le2\theta.
\label{eq:angle}
\end{align}
The associated cosine interval is clipped to $[0,\pi/2]$ in angle. No
retained-system gap or norm appears in these bounds.
\end{theorem}
\begin{proof}
Let $H_{d,z}=H_z+\sum_i a_i X_{A_i}$ and $e_{d,z}$ be its ground energy.
It commutes with $P_z$. At every fixed core string, one excited driven leaf
costs at least $\Delta_0$. Thus
$Q_zH_{d,z}Q_z\ge(e_{d,z}+\Delta_0)Q_z$.
A ground vector of $H_{d,z}$ can be chosen with a fixed driven-core string;
its expectation of each $X_{A_i}$ vanishes. The variational principle gives
$E_z\le e_{d,z}$. Since $\|\sum_i a_iX_{A_i}\|=r$,
\begin{equation}
Q_zH_zQ_z-E_zQ_z\ge\delta Q_z.
\label{eq:highblock}
\end{equation}

Only driven-core flips connect $P_z$ and $Q_z$. A flip at cell $i$ changes
its conditional leaf vacuum with overlap $R_i$. Its excited component has
squared norm $1-R_i^2$, independent of the core string and host state.
Excited components produced at different cells are orthogonal, since only
the leaves of the flipped cell are excited. Consequently the operator
identity, not just a norm inequality, is
\begin{equation}
P_zH_zQ_zH_zP_z=b_*^2P_z.
\label{eq:offblock}
\end{equation}
Using Eq.~\eqref{eq:highblock} in the $Q_z$ ground-state equation yields
$\|Q_z\Omega_z\|\le(b_*/\delta)\|P_z\Omega_z\|$, proving
Eq.~\eqref{eq:leakage}. Each full state is at Fubini--Study distance at
most $\theta$ from its normalized projection. Those projections have
amplitude overlap $cF_T$ by Eq.~\eqref{eq:scalar}. The triangle inequality
for projective distance proves Eq.~\eqref{eq:angle}.
\end{proof}
One immediate local-only ceiling is
\begin{equation}
\Fglobal\le\cos\bigl([\arccos c-2\theta]_+\bigr).
\end{equation}
It can place the bare overlap below $F^2=1/2$, the sufficient threshold
for a nonzero Ramsey amplitude at all times derived in
Eq.~\eqref{eq:ramsey}, even when the retained states have overlap almost
one. Below this threshold, that bound alone does not determine whether
the amplitude reaches zero.

\begin{figure}[ht]
\centering
\includegraphics[width=.65\textwidth]{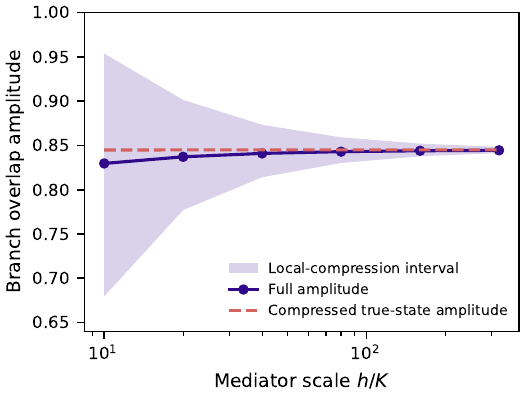}
\caption{Evaluation of Eq.~\eqref{eq:angle} for the full four-cell model.
Here $j/h=1$, $g/h=0.55$, $K=1$, and the absolute mismatch $j\epsilon=0.2$
is fixed; hence the relative mismatch changes with $h$. The nominal projected
bulk field is $a=\sqrt{1+(j_+/h)^2}\sqrt{1+(j_-/h)^2}$, not an independently
located critical field. Shading is the analytic interval evaluated on
numerically computed true retained states. It describes the compression
error for those states, rather than statistical uncertainty or an error
estimate for a projected-Hamiltonian ground state.}
\label{fig:finite-mediator}
\end{figure}

Figure~\ref{fig:finite-mediator} evaluates the local angle bound on the
full four-cell model as the mediator energy is increased.

\section{A local decoder and a precisely defined Ramsey protocol}
Extend each spinor to the real rotation
\begin{equation}
U(f)=\begin{pmatrix}n_1(f)&-n_2(f)\\n_2(f)&n_1(f)\end{pmatrix}.
\end{equation}
Let $\mathcal E_0$ attach the reference state $|Z=+1\rangle$ to every
driven leaf while leaving the retained system unchanged. The product
rotation $U_z$, conditioned on each driven core orientation $Z_{A_i}$,
uses the two local fields specified in Sec.~\ref{sec:compression}.
It maps $\mathcal E_0$ to $\mathcal E_z=U_z\mathcal E_0$.
Applying $U_z^\dagger$ therefore aligns the conditional leaf states with
the same reference state in both branches. The decoded full states are
\begin{equation}
\widetilde\Omega_z=U_z^\dagger\Omega_z
=\sqrt{1-q_z}\,\mathcal E_0\psi_z+e_z,
\quad \mathcal E_0^\dagger e_z=0,\quad\|e_z\|^2=q_z.
\end{equation}
Writing $F_{\rm dec}=|\langle\widetilde\Omega_-|\widetilde\Omega_+\rangle|$
and using Cauchy--Schwarz gives
\begin{align}
|F_{\rm dec}-F_T|
&\le1-\sqrt{(1-q_-)(1-q_+)}+\sqrt{q_-q_+}\nonumber\\
&\le2\sin^2\theta.
\label{eq:decode}
\end{align}
This is an absolute amplitude estimate of order $(a/\Delta_0)^2$ under
large-local-gap scaling. The dressed qubit transition is
$\mathcal K=U_-U_+^\dagger$. It acts on driven cores and their leaves; it is not an
operation only on the leaves. The erasure theorem applies to its full finite
support when that support satisfies the common-basis positivity hypotheses.

For a specified unitary $\mathcal K$, prepare the environment in the $+$ branch
ground state and perform a qubit pulse that applies $\mathcal K$ on the minus branch.
The closing pulse applies the inverse conditional operation, so the
branches recombine in the same rotated basis. The amplitude is then
$\ell_{\mathcal K}(t)=\langle\Omega_+|\mathcal K^\dagger e^{iH_-t}
\mathcal K e^{-iH_+t}|\Omega_+\rangle$. The same matrix element enters
both preparation and recombination, giving nonnegative spectral weights:
\begin{equation}
\ell_{\mathcal K}(t)=e^{-iE_{0,+}t}
\sum_n|\langle n,-|\mathcal K|\Omega_+\rangle|^2e^{iE_{n,-}t}.
\end{equation}
The total weight is one. The ground term has weight
$Z_{\mathcal K}=|\langle\Omega_-|\mathcal K|\Omega_+\rangle|^2$, or the sum over the ground
space when degenerate. Therefore
\begin{equation}
|\ell_{\mathcal K}(t)|\ge\max\{0,2Z_{\mathcal K}-1\}.
\label{eq:ramsey}
\end{equation}
No equality of branch ground energies is required. A full trace-norm initial
state error $d$ changes the expectation of the unitary interferometric
operator by at most $d$. This proves the modified bound
$\max\{0,2Z_{\mathcal K}-1-d\}$. It does not account for preparation time or imperfect
implementation of $\mathcal K$. It differs from a sudden quench from a reference
Hamiltonian's ground state and from a Gibbs-state protocol.

\section{The separate cost of using a projected-Hamiltonian ground state}
Let $A_z=\mathcal E_z^\dagger H_z\mathcal E_z$ have nondegenerate ground state $\phi_z$ and
gap $\gamma_z>0$. Eliminating $Q_z\Omega_z$ gives exactly
\begin{equation}
[A_z-\Sigma_z(E_z)]\psi_z=E_z\psi_z,
\quad
0\preceq\Sigma_z(E_z)\preceq\eta I,
\quad \eta=b_*^2/\delta.
\end{equation}
This is the standard block-resolvent/Feshbach step underlying effective
Hamiltonian constructions \cite{Bravyi2011}. Since $E_z\le\lambda_{\min}(A_z)$,
projection of this equation onto the excited subspace of $A_z$ yields
\begin{equation}
\sin\angle(\psi_z,\phi_z)\le\min\{1,\eta/\gamma_z\}.
\label{eq:retainedgap}
\end{equation}
Consequently replacing $F_T$ in Eq.~\eqref{eq:angle} by
$|\langle\phi_-|\phi_+\rangle|$ adds at most
$\sum_z\arcsin[\min(1,\eta/\gamma_z)]$ to its angle budget. This additional
step can fail to be informative when the retained gap closes. A numerical
Ritz gap is not a certified lower gap bound, so intervals evaluated using
floating-point gaps are diagnostics, not rigorous eigenvalue enclosures.

The logical distinction is exposed by the exact matrix
\begin{equation}
H=\begin{pmatrix}0&0&0\\0&v^2/(2\Delta)&v\\0&v&\Delta\end{pmatrix},
\qquad P=\operatorname{diag}(1,1,0).
\label{eq:counter}
\end{equation}
The lower eigenvalue of the last two levels is negative, so the full ground
state compresses onto the second retained vector. The ground state of $PHP$
is the first retained vector. Their overlap is zero, even as the discarded
weight tends to zero with $v/\Delta$. This is a counterexample to an
inference about local leakage, not a claimed phase of the pendant chain.

\section{Independent calculations and numerical controls}
The full pendant solver builds the original Pauli Hamiltonian in the product
$Z$ basis. There are $L$ three-spin cells; only the four leaves of driven cells
0 and 1 are compressed. An independent construction gives $A_z$ and checks
it against $\mathcal E_z^\dagger H_z\mathcal E_z$. Each branch is solved separately. Complete
source susceptibilities, where reported, use the entire projected resolvent
$(H-E_0)^{-1}$ on the ground-state orthogonal complement, not a first pole.
No critical field is fitted or certified.

Root fidelity after erasing selected spins is evaluated without constructing
a large retained density matrix. If the full state amplitude matrices are
$A_{xy},B_{xy}$, then
\begin{equation}
F_R=\|A^\dagger B\|_1,
\qquad F_Z=\sum_x\|A_{x\cdot}\|\,\|B_{x\cdot}\|.
\end{equation}
This small singular-value calculation follows directly from Uhlmann's
formula. A separate test compares it to explicit density-matrix square roots.

The finite-chain calculations cover 27
parameter points: 18 source scans at $L=4$, six local-energy controls at
$L=4$, and three extra sizes $L=3,5,6$. The largest full Hilbert space is
$2^{18}=262144$; the corresponding exact retained space has dimension
$2^{14}=16384$. Sixty-four complex conditional-host Hamiltonians check
Theorem~\ref{thm:compression}. Ninety-six additional real Hamiltonians with
nonpositive off-diagonal entries and arbitrary diagonal selected-spin
interactions check Theorem~\ref{thm:erasure}. Full states, parameters,
and local amplitude-ratio diagnostics accompany the numerical data.
Amplitude-ratio checks exclude unresolved coefficients below $10^{-13}$ and
report their probability mass; they are not pointwise interval proofs.

\begin{table}[ht]
\centering
\caption{Representative complete four-cell results with $K=1$, $h=40$,
$j/h=0.5$, $\epsilon=0.01$ and the stated nominal projected bulk field.
$Z_R$ is the optimal transition weight after erasing the four driven leaves.}
\begin{tabular}{cccc}
\toprule
$g/h$ & $Z_{\rm full}$ & $Z_R$ & $Z_{\rm dec}$\\
\midrule
0.55 & 0.461413691 & 0.994699296 & 0.994688505\\
0.90 & 0.143543322 & 0.987206288 & 0.987197439\\
1.20 & 0.043791769 & 0.980992399 & 0.980987887\\
\bottomrule
\end{tabular}
\label{tab:data}
\end{table}
All table entries are rounded; full precision is retained in the numerical
data. These large differences between bare and locally decoded weights are
finite local-record effects, not a fit of an orthogonality exponent.

\begin{figure}[ht]
\centering
\includegraphics[width=.65\textwidth]{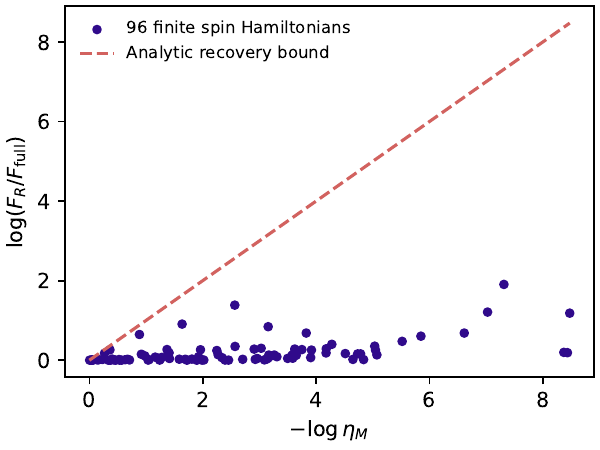}
\caption{The logarithmic fidelity loss caused by reinstating the erased
region lies below its proven local bound for 96 finite Hamiltonians.
The line is an analytic bound, not a fit. The finite tests do not establish
thermodynamic scaling or global sharpness; sharpness is proved by the product
example in Sec.~3.1.}
\end{figure}

The maximum full-ground residual is $3.31\times10^{-11}$ and the maximum
complete-response linear-system residual is $2.82\times10^{-11}$. These are
floating-point diagnostics, not certified error bars. Independent checks
compare the small-matrix and density-matrix fidelity formulas, verify decoder
unitarity, and compare complete Pauli and Lehmann response calculations.
Spectral gaps are obtained by full diagonalization for small matrices and
symmetry-mixed Lanczos starts for larger ones, so that the two lowest states
need not belong to the same symmetry sector. Overlaps exceeding one by
normalization roundoff of order $10^{-15}$ are clipped to one for reported
weights; the unmodified values are retained in the numerical data.

\section{Relation to established results and scope}
Ordinary fidelity monotonicity and purification optimization follow from
the established fidelity theory \cite{Uhlmann1976,Jozsa1994}; optimal
coherence recovery with limited environmental access has also been studied
\cite{Miatto2015}.
Theorem~\ref{thm:erasure} supplies the reverse comparison with explicit
local constants for the stated Hamiltonian class, leading to invariant
orthogonality exponents and a necessary control cost. The scalar
$W_\epsilon^4$ factor follows from the conditional embedding in
Eq.~\eqref{eq:scalar}; Theorem~\ref{thm:compression} bounds the difference
between this embedded description and the full finite-mediator ground states.

Local stability theorems also include situations where a global gap can
close \cite{Henheik2022}. They control local observables under perturbations
with specified local-gap and weak-interaction structure.
Theorem~\ref{thm:erasure} compares
fidelity before and after discarding a finite subsystem, using common-basis
positivity and bounded local longitudinal biases rather than a gap. Energy-penalty
leakage results \cite{Marvian2015} address another related but distinct task.

The connected sign-defect chain in Sec.~\ref{sec:critical-example} supplies
an exact thermodynamic orthogonality exponent and its optimal local-control
consequence. Phase boundaries of the finite-mediator model and
arbitrary-initial-state, finite-temperature, or long-time Ramsey decay laws
are separate questions. The compression theorem controls the true retained
state; replacing it by a projected-Hamiltonian ground state requires the
additional gap estimate in Eq.~\eqref{eq:retainedgap}.

\begingroup\footnotesize
\bibliographystyle{apsrev4-2}
\bibliography{references}
\endgroup